\documentclass[12pt]{article}
\usepackage{amssymb}
\usepackage{graphicx}
\usepackage{amsmath}
\usepackage{amsfonts}
\usepackage{hyper}
\usepackage{float}

\begin{document}

\title{Conformal Affine Toda Soliton and Moduli of IIB Superstring on $%
AdS_5\times S^5$}
\author{\hspace{5mm}Bo-Yu Hou$^\diamondsuit$\thanks{%
Email:byhou@nwu.edu.cn}, Bo-Yuan Hou$^\P$, Xiao-Hui Wang$^\diamondsuit$%
\thanks{
Email:xhwang@nwu.edu.cn,} \\
Chuan-Hua Xiong$^{\diamondsuit }$\thanks{
Email:chhxiong@nwu.edu.cn,}, Rui-Hong Yue$^{\diamondsuit }$\thanks{%
Email:rhyue@nwu.edu.cn}\\
$^\diamondsuit$\textsl{Institute of Modern Physics, Northwest University,}\\
\textsl{Xi'an, 710069, P. R. China}\\
$\P $\textsl{Graduate School, Chinese Academy of Science,}\\
\textsl{\ Beijing 100039, P. R. China}}
\date{}
\maketitle

\centerline{Abstract} \bigskip In this paper we construct the $%
gl(2,2|4)^{(1)}$ affine gauged WZW action of type IIB Green-Schwarz
superstring in the $AdS_{5}\times S^{5}$ coset space. From this gauged WZW
action we obtain the super Lax connection. Further, in terms of torsion
relation we get the pure Bose Lax connection and argue that the IIB
Green-Schwarz superstring embedded in $AdS_{5}\times S^{5}$ is the $\mathbb{%
CP}^{3}$ conformal affine Toda model. We review how the position $\mu $ of
poles in the Riemann-Hilbert formulation of dressing transformation and how
the value of loop parameters $\mu $ in the vertex operator of affine algebra
determines the moduli space of the soliton solutions, which describes the
moduli space of the Green-Schwarz superstring. We show also how this affine $%
SU(4)$ symmetry affinize the conformal symmetry in the twistor space, and
how a soliton string corresponds to a Robinson congruence with twist and
dilation spin coefficients $\mu $ of twistor.

\newpage

\section{Introduction}

The AdS/CFT duality \cite{mal} set up the relations between the bulk
classical theory and the boundary quantum theory. There exist many hints of
the integrable structure on the both sides. Bena, Polchinski and Roiban \cite%
{Pol,polykov} find the infinite set of nonlocal conserved currents that is
the hidden symmetry for the IIB Green-Schwarz superstring in $AdS_{5}\times
S^{5}$. Dolan \cite{witten,dw04} et. al. study the relation with quantum
integrability in $D=4,\mathcal{N}=4$ super Yang-Mills theory.

However, all the suggested integrable models of both sides was given after
some approximation \cite{classical,quantum}. In the classical bulk, the
exact rotating string in $AdS_{5}\times S^{5}$ after taking harmonic
expansion around point string has been described by (confer the last
reference in Ref.\cite{classical}) a Neumann-Rosochatius one-dimensional
integrable system \cite{neumann1,neunann2}. For the quantum boundary CFT,
one use the Feynmann diagram of tree level and one-loop to relate it with
Bethe Ansatz of some spin chain \cite{min,bei,mar,bel}. Then the classical
and quantum anomalous dimensions and momentum are compared order by order.

\textbf{Symmetry dictates interactions }\cite{cnyang}. Metsaev and Tseylin
\cite{mt1} find that the global $PSU(2,2|4)$ super-invariance and local $%
\kappa $ symmetry dictates the unique IIB string action on $AdS_{5}\times
S^{5}$. Roiban Siegel \cite{RS} bring the coset to $GL(4|4)/(sp(4)\otimes
GL(1))^{2}$ in the more transparent $\kappa $ symmetric gauge, give the
simplest action with manifest conformal symmetry.

In this paper, our aim is to argue that the hidden symmetry has to dictate
the classical action with the symmetry broken by anomaly, by the vacuum
expectation value of chiral field. The $GL(1)^{2}\ $transformation of the
complex parameter $\mu $, which labels the vacua (the moduli space),
realizes as the opposite reparametrization of left and right moving string.
This loop group is further central extended by the 2 cocycle of WZW term. In
summary, the hidden affine symmetry $gl(4,4)^{(1)}$ should dictate a gauged
WZW action. After shortly review Roiban-Siegel action \cite{RS}, we suggest
the axially gauged WZW action for super current with constraint label by $%
\mu ,$ then sketch the derivation of the pure bosonic EOM and its
corresponding gauged WZW action. This EOM and action describe the dynamics
of the chiral embedding of IIB string. The left and right evolution of
moving frame is expressed as the Lax connection, which is uniquely
determined by fields of diagonal Cartan element. It turns out to be the
conformal affine Toda \cite{b244,babelon,underwood}.

Once the bosonic EOM and (or) the gauged WZW action has been obtained, it is
already obvious that it should dictate the conformal affine Toda, and to
ascertain that the exact integrable model of both classical and quantum
AdS/CFT are the affine Toda. In section 4, we only review the key and subtle
point for the chain: gauged WZW to chiral embedding to conformal affine
Toda, then review its soliton solution in section 5. \ All this is well
know, so the readers better skip the section 4 and 5, go straightforward to
the solitonic string picture and turn back in case of need. If something is
unclear, then this part may serve as a directory\ to original papers. But we
would like to stress: 1. The Riemann-Hilbert problem of dressing symmetry,
description of the holomorphic and anti holomorphic behavior of left and
right moving deserves\ attention. We will discuss the Riemann-Hilbert
factorization in the next paper. 2. The vertex construction is not only
effective to construct classical solution \cite{b244,babelon,underwood}, but
also implies the exact quantum version. 3. The geometry of pseudo-sphere
\cite{chen,hhbook,houo3} in connection with nonlinear $\sigma $ model, with
sine-Gordon, with Backlund transformation and R.H. lies in the heart of
duality and dressing twist, but such kind of geometry seems not popular till
now.

At last we find the \textbf{ground states} are \textbf{stretched string}.
Its radius of deformation and correspondent energy-momentum are
characterized by moduli parameter $\mu .$ The hidden symmetry extend the
conformal symmetry describes by twistor \cite{penrose}, to that of a
\textquotedblleft affine\textquotedblright\ conformal symmetry of twistor,
the stretched string realize the Robinson congruence \cite{penrose} of ray
\cite{prani,sachs}.

\section{Roiban-Siegel's super vielbein construction and \newline
quadratic WZW term of IIB string on $AdS_{5}\times S^{5}$}

The Metsaev-Tseytlin action \cite{mt1} of IIB Green-Schwarz superstring in $%
AdS_{5}\times S^{5}$ is realized as the sigma model on coset $\frac{%
PSU(2,2|4)}{SO(4,1)\times SO(5)}$.

Roiban and Siegel \cite{RS} extend the symmetry into $\mathbf{GL(2,2|4)}$ by
adding two $\mathbf{U(1)}$ factors
\begin{equation}
\frac{PSU(2,2|4)}{SO(4,1)\times SO(5)}\longrightarrow \frac{GL(2,2|4)}{%
SO(4,1)\times SO(5)\otimes U(1)^{2}}.
\end{equation}%
Further, using a Wick rotation, it gives
\begin{equation}
\frac{GL(2,2|4)}{SO(4,1)\times SO(5)\times U(1)^{2}}\longrightarrow \frac{%
GL(4|4)}{(SP(4)\times U(1))^{2}}.
\end{equation}

The coset representative vielbeins $Z_{M}^{\ B}$ is given by
\begin{equation}
Z_{M}^{\ B}=\Big[x^{(4)}\Big]\Big[\theta \Big]\Big[x^{(6)}\Big]=\left(
\begin{array}{cc}
X_{m}^{\ d} & 0 \\
0 & \delta _{\bar{m}}^{\ \bar{n}}%
\end{array}%
\right) \left(
\begin{array}{cc}
\delta _{d}^{\ c} & \theta _{d}^{\ \bar{p}} \\
\theta _{\bar{n}}^{\ c} & \delta _{\bar{n}}^{\ \bar{p}}%
\end{array}%
\right) \left(
\begin{array}{cc}
(X_{0})_{c}^{\ b} & 0 \\
0 & z_{\bar{p}}^{\ \bar{b}}%
\end{array}%
\right) \ .  \label{vierbein}
\end{equation}

The indices are introduced by

\begin{center}
$%
\begin{array}{cccl}
GL\left( 4\right) ^{2} & N=(m,\bar{m}) &
\begin{array}{c}
\mathrm{even}GL(4) \\
\mathrm{odd}GL(4)%
\end{array}
&
\begin{array}{c}
m,n=1,\cdots 4 \\
\bar{m},\bar{n}=1,\cdots 4%
\end{array}
\\
Sp(4)^{2} & A=(a,\bar{a}) &
\begin{array}{c}
\mathrm{even}Sp(4) \\
\mathrm{odd}Sp(4)%
\end{array}
&
\begin{array}{l}
a,b=1,\cdots 4\quad \mathrm{further},a=(\alpha ,\dot{\alpha}),\alpha =1,2.
\\
\bar{a},\bar{b}=1,\cdots 4%
\end{array}%
\end{array}%
$
\end{center}

The currents are given by
\begin{equation}
J_{N}^{\ M}+A_{N}^{\ M}=(Z^{-1})_{A}^{\ M}dZ_{N}^{\ A},  \label{ctf}
\end{equation}%
where $A_{N}^{\ M}$ are the $\left( Sp(4)\times GL(1)\right) ^{2}$
connection.

The antisymmetrization, tracelessness, index contraction and inverse of $%
\Omega $ are defined as\cite{RS}
\begin{equation}
\begin{array}{rl}
A_{[m}B_{n]}=\frac{1}{2}\left( A_{m}B_{n}-A_{n}B_{m}\right) \ , &
A_{<m}B_{n>}=A_{[m}B_{n]}+\frac{1}{4}\Omega _{mn}A^{P}B_{P}\ , \\[3mm]
A^{m}B_{m}=\Omega ^{mn}A_{m}B_{n}\ , & \Omega ^{mn}\Omega _{nP}=\delta
_{P}^{\ m}.%
\end{array}%
\end{equation}%
Here $\mathrm{\Omega =diag(\omega ,\omega )}$ and $\omega $ is the
symplectic metric of $Sp(4)$. Only the antisymmetrical traceless current $%
J_{N}^{\ M}$ is \textbf{dynamical}. Thus the kinetic part is
\begin{equation}
S_{k}=\int J^{<mn>}\wedge \ast J_{<mn>}-J^{<\bar{m}\bar{n}>}\wedge \ast J_{<%
\bar{m}\bar{n}>}.  \label{sk}
\end{equation}

To keep the $\kappa $\textbf{\ symmetry}, Roiban and Siegel find the
topological $S_{wzw}$ should be
\begin{equation}
S_{wzw}=\pm \int \frac{1}{2}\left( E^{1/2}J^{m\bar{n}}\wedge J_{m\bar{n}%
}-E^{-1/2}J^{\bar{m}n}\wedge J_{\bar{m}n}\right) ,
\end{equation}%
where $E=\mathrm{sdetZ_{N}^{\ M}}$, this sign $\pm $ is used instead of $%
S^{IJ}=\mathrm{diag}(1,-1)(I=1,2)$ in Ref.\cite{mt1} and for the $\theta _{%
\bar{m}}^{Im},$ $\kappa ^{I}$ etc..

The total action is
\begin{equation}  \label{eq8}
S=S_k+S_{wzw}.
\end{equation}

The simplest action is given by choosing the $\kappa $ symmetric gauge
\begin{equation}
\theta _{\alpha }^{\ \bar{n}}=0\ ,\quad \theta _{\bar{m}}^{\ \dot{\alpha}}=0.
\end{equation}

In this gauge, only 16 $\theta _{\dot{\alpha}}^{\ \bar{n}},\theta _{\bar{m}%
}^{\ \alpha }$ survives. The explicit form of the currents is given by
\begin{equation}
\begin{array}{rcl}
J_{a}^{\ b} & = & (j_{ads_{5}})_{a}^{\ b}-(X_{0}^{-1})_{a}^{\ c}\theta
_{c}^{\ \bar{m}}d\theta _{\bar{m}}^{\ d}(X_{0})_{d}^{\ b}\ , \\[3mm]
J_{\bar{a}}^{\ \bar{b}} & = & (j_{S^{5}})_{\bar{a}}^{\ \bar{b}}=(z^{-1})_{%
\bar{a}}^{\ \bar{m}}dz_{\bar{m}}^{\ \bar{b}}\ , \\[3mm]
J_{a}^{\ \bar{b}} & = & (X_{0}^{-1})_{a}^{\ c}d\theta _{c}^{\ \bar{m}}z_{%
\bar{m}}^{\ \bar{b}}\ , \\[3mm]
J_{\bar{a}}^{\ b} & = & z_{\bar{a}}^{\ \bar{m}}d\theta _{\bar{m}}^{\
c}(X_{0})_{c}^{\ b}\ ,%
\end{array}%
\end{equation}

where
\begin{equation}
j_{ads_{5}}=\left(
\begin{array}{cc}
-\frac{dx^{0}}{x^{0}}\omega & \frac{1}{x^{0}}d\mathbf{x}^{T}\omega \\
\frac{1}{x^{0}}\omega d\mathbf{x} & \frac{dx^{0}}{x^{0}}\omega%
\end{array}%
\right) .  \label{11}
\end{equation}

\section{The Gauged WZW Action of IIB string on $AdS_{5}\times S^{5}$}

The $gl(4,4)$ symmetry and $\kappa $ symmetry have dictated Roiban Siegel's
action. Now the \textbf{reparametrization symmetry} will dictate \textbf{%
gauged WZW action}.

\subsection{Fermionic action}

In order to obtain the spontaneously broken ground state of the IIB string
in $AdS_{5}\times S^{5}$, we will add the D and FI terms which are
determined by the vacuum expectation values. Thus we must gauging the WZW
action (\ref{eq8}) i.e. find the Hamiltonian reduction of (\ref{eq8})%
\footnote{%
Remark: we choose the RS action instead of the beautiful Berkovits formalism
of superstring quantization on $AdS_{5}$ background\cite{berkovits0001035},
since there the $\kappa $ symmetry is not manifest. The kinetic part $%
\langle \left( g^{-1}\partial _{i}g\right) ^{2}\rangle $\ includes the
Cartan form of Lorentz generators \cite{9902098,9907200}, thus no dressing
symmetry. The WZW is gauged by vector gauge \cite{9907200}, not the axial
one, thus the chiral symmetry seems not manifest also.}. The IIB string is
chirally embedded \cite{gervais} in $AdS_{5}\times S^{5}$, such that the
left and right moving tangent vectors of the world sheet is mapped to the
two co-moving tangents of $AdS_{5}\times S^{5}$ respectively. But meanwhile,
by SUSY, the $N=2$ fermionic beins $L^{I}(I=1,2)$ should be rotated by axial
$U(1)$\ in accompany. The Cartan forms which describes the string embedded
in the $AdS_{5}$ should be corresponded to the $E_{+},E_{-}$ components of a
$sl(2)$ subalgebra in $gl(2,2|4)$ and this $E_{+},E_{-}$ and axial $U(1)$\
are generated by the primary generator $F_{\alpha },F_{\dot{\alpha}}\left(
\alpha =1,2\right) $ of the $Osp(2|2).$Here the $N=1$ $osp(2|1)$ embedding
obviously is excluded. The principal embedding of $sl(2|1)$ in fundamental
representation can not give the decoupled diagonal bosonic block with a $%
SL(2)$ subgroup in upper left block for $AdS_{5}$ and in lower right block
the extended supersymmetry axial $U(1).$ Only the adjoint $Osp(2,2)$
embedding works this way.

The super covariant derivatives on $AdS_{5}\times S^{5}$\ are
\begin{equation}
D_{n}^{\ \bar{m}}=\frac{\partial }{\partial \theta _{\bar{m}}^{\ n}}+\theta
_{p}^{\ \bar{m}}\left( \gamma ^{m}\right) _{n}^{\ p}\frac{\partial }{%
\partial x^{m}}\ ,
\end{equation}%
\begin{equation}
D_{\bar{m}}^{\ n}=\frac{\partial }{\partial \theta _{n}^{\ \bar{m}}}+\theta
_{\bar{m}}^{\ p}\left( \gamma ^{m}\right) _{p}^{\ n}\frac{\partial }{%
\partial x^{m}}\ ,
\end{equation}%
where the target $\theta $ and x are functions $\theta (\tau ,\sigma ),%
\mathrm{x}(\tau ,\sigma )$ of the world sheet variable $\tau ,\sigma $. In
the super covariant symmetrical killing gauge
\begin{equation}
\theta _{\alpha }^{\ \bar{n}}\left( x,\theta \right) =0\ ,\quad \theta _{%
\bar{m}}^{\ \dot{\alpha}}\left( x,\theta \right) =0\ .  \label{kappa}
\end{equation}%
Surviving $D_{n}^{\bar{m}},D_{\bar{m}}^{n},$becomes
\begin{equation}
D_{\alpha }^{\ \bar{m}}=\frac{\partial }{\partial \theta _{\bar{m}}^{\
\alpha }}+\theta _{\dot{\alpha}}^{\ \bar{m}}\left( \gamma ^{m}\right)
_{\alpha }^{\ \dot{\alpha}}\frac{\partial }{\partial x^{m}}
\end{equation}%
\begin{equation}
D_{\bar{m}}^{\ \dot{\alpha}}=\frac{\partial }{\partial \theta _{\dot{\alpha}%
}^{\ \bar{m}}}+\theta _{\bar{m}}^{\ \alpha }\left( \gamma ^{n}\right)
_{\alpha }^{\ \dot{\alpha}}\frac{\partial }{\partial x^{n}}.  \label{16}
\end{equation}%
They commute with the supercharge $Q_{\alpha },Q_{\dot{\alpha}}.$

The super currents is defined similar with (\ref{ctf})
\begin{equation}
Z^{-1}D_{\alpha }Z=\mathcal{J}_{\alpha }
\end{equation}%
here%
\begin{equation*}
{\mathcal{J}}_{\alpha }={\mathcal{J}}_{\alpha }^{MN}T_{MN},
\end{equation*}%
we adapt similar notation for $\mbox{\boldmath$\nu$}_{\alpha },%
\mbox{\boldmath$\mu$}_{\alpha }$\ later. But now the even derivative $d$ on
the world sheet is replaced by the odd $D_{\alpha }$ and $D_{\dot{\alpha}}$
pulled back by the static map $\left( \sigma _{i}\longrightarrow x,\theta
\right) $ to the world-sheet.

The embedding in $gl(4,4)^{(1)}$ is given by the $Osp(2|2)$ generators
\begin{equation}
T_{\alpha }=\sum_{\bar{m}}T_{\alpha }^{\ \bar{m}},\quad T^{\dot{\alpha}%
}=\sum_{\bar{m}}T_{\bar{m}}^{\ \dot{\alpha}},
\end{equation}%
\begin{equation}
\left\{ T^{1},T^{2}\right\} =E^{+},\{T^{\dot{1}},T^{\dot{2}}\}=E^{-}.
\end{equation}%
Remark: The $T_{\alpha }^{\ \bar{m}}\left( T_{\bar{m}}^{\ \dot{\alpha}%
}\right) $ lies not in the upper right (lower left) off diagonal block as
usually (e.g. \cite{mtsu4} appendix) in distinguished basis. It is, as in
WZW term (\ref{eq8}) of RS, the antisymmetric combination of conjugate terms
in upper right and lower left, and has different phase for $I=(1,2)$
components in the $SO(2)$ of $N=2$ in \cite{mt1}.

\subsection{Bosonic EOM and gauged WZW action of IIB string in $AdS_{5}$}

In the Killing gauge, we find the pure bosonic integrable condition\footnote{%
We will give it in the next paper.}:
\begin{equation}
\partial _{\pm }U=A_{\pm }U.  \label{au}
\end{equation}

Using the $U$ one can obtain the pure bosonic gauged WZW action
\begin{eqnarray}
I\left( U,\mathbb{A}_{+},\mathbb{A}_{-}\right) &=&S(U)+\kappa \int d^{2}x{%
\mathrm{T}r\{}\mathbb{A}_{-}\left( \partial _{+}U\right) U^{-1}+\left(
U^{-1}\partial _{-}U\right) \mathbb{A}_{+}  \notag \\
&&+\mathbb{A}_{-}U\mathbb{A}_{+}U^{-1}-\mathbb{A}_{-}\mbox{\boldmath$\mu$}-%
\mathbb{A}_{+}\mbox{\boldmath$\nu$}\}.  \label{acction}
\end{eqnarray}

The EOM of pure bosonic can be derived from this gauged WZW\cite{belog}%
\begin{equation}
\partial _{+}(z^{-1}\partial _{-}z+z^{-1}\mathbb{A}_{-}z)-[\mathbb{A}%
_{+,}z^{-1}\partial _{-}z+z^{-1}\mathbb{A}_{-}z]+\partial _{-}\,\mathbb{A}%
_{+}=0\ ,  \label{em1}
\end{equation}%
\begin{equation}
\partial _{-}(\partial _{+}zz^{-1}+z\mathbb{A}_{+}z^{-1})+[\mathbb{A}%
_{-,}\partial _{+}zz^{-1}+z\mathbb{A}_{+}z^{-1}]+\partial _{+}\,\mathbb{A}%
_{-}=0\ ,  \label{em2}
\end{equation}%
\begin{equation}
{\mathrm{T}r}[E_{+}(z^{-1}\partial _{-}z+z^{-1}\mathbb{A}_{-}z-%
\mbox{\boldmath$\nu$})]=0,\quad \mbox{\boldmath$\nu$}\equiv \mu _{-}E_{-}\ ,
\label{em3}
\end{equation}%
\begin{equation}
{\mathrm{T}r}[E_{-}(\partial _{+}zz^{-1}+z\mathbb{A}_{+}z^{-1}-%
\mbox{\boldmath$\mu$})]=0,\quad \mbox{\boldmath$\mu$}\equiv \mu _{+}E_{+}\ .
\label{em4}
\end{equation}

One can show that for $AdS_{5}$\ coset $SU(4)/SP(4)$ the vierbein $z$ and
the nonlinear representative G can be identified in some gauge. So here and
latter, we will use the $U^{-1}dU$ (i.e. the $g^{-1}dg$ in \cite{belog}, the
$T^{-1}dT$ in \cite{b244}) instead of $z^{-1}dz$.

It is well known \cite{belog,6,7,8} that gauged WZW model given by the
action (\ref{acction}) or EOM(\ref{em1}-\ref{em4}) are equivalent to the
chiral embedding (Gervais and Matsuo \cite{gervais}). The Green-Schwarz IIB
superstring propagate on the $\mathrm{AdS_{5}\times S^{5}}$ background can
be regard as the following chiral embedding . The IIB superstring embed in
only the $\mathrm{AdS_{5}}$ of the $\mathrm{AdS_{5}\times S^{5}}$. Consider
the image on tangent plane of $AdS_{5}$ of the two left and right moving
tangent vectors respectively of world-sheet as the moving frame. The
rotation of the moving frame is given by the Frenet-Serret equations \cite%
{gervais} in the $\mathbb{CP}^{3}$ manifold (actually the vector
representation space of affine $SL(4)$ if we include the twisted dilation
and phase variation of $\mu ,\nu $).\

\section{Chiral embedding and conformal affine Toda model}

The chiral embedding dictates the Toda model \cite{gervais}. Later these has
been generalized to the affine case \cite{underwood}. In this paper we will
apply these well known result for the chiral embedding of IIB string in $%
AdS_{5}$ to find stretched{\huge \ }string soliton.

At first let us formulate the conclusion. The U in EOM (\ref{em1}-\ref{em4})
and action (\ref{acction}) is the \textbf{transfer matrix} of \textbf{%
conformal affine Toda} model written in principal curvature coordinate \cite%
{houo3},\cite{b244} as
\begin{equation}
\partial _{\pm }U=A_{\pm }U.  \label{au}
\end{equation}%
But, here the Lax connection $A_{\pm }$ is in different gauge with $\mathbb{A%
}_{\pm }$ in (\ref{acction}-\ref{em4}),%
\begin{eqnarray}
A_{+} &=&\mathbb{A}_{+}^{\Phi }\equiv e^{\Phi }\mathbb{A}_{+}e^{-\Phi
}+e^{\Phi }\mathbb{\partial }_{+}e^{-\Phi },  \notag \\
A_{-} &=&\mathbb{A}_{-}^{-\Phi }\equiv e^{-\Phi }\mathbb{A}_{-}e^{\Phi
}+e^{-\Phi }\mathbb{\partial }_{-}e^{\Phi }.  \label{a}
\end{eqnarray}%
Here $\Phi $ is the affine Toda field (\ref{38}).

Now we turn to \textbf{chiral embedding}, to obtain the conformal Affine
Toda $\Phi $ after (\ref{38}).

Let $U_{L},U_{R}$\ be the rotation of moving frame of left and right chiral
embedding \cite{gervais}. We will show that they are the transfer matrix $%
U_{L},U_{R}$ (the $g_{1},g_{2}$ in \cite{b244}) in the triangular gauge or
in the asymptotic line gauge. The transfer matrices \textquotedblleft
U\textquotedblright\ in different gauge satisfy%
\begin{equation}
U_{L}=e^{\Phi }U,U_{R}=e^{-\Phi }U.  \label{uu}
\end{equation}%
where the element $\Phi ,$ will be given later in (\ref{38}).

Now let's gauss decompose $U_{L},U_{R},$
\begin{equation}
U_{L}\equiv e^{K_{-}}N_{+}M_{-},\quad U_{R}\equiv e^{K_{+}}N_{-}M_{+}.
\label{u}
\end{equation}%
Here $K_{-},K_{+}$ is the diagonal Cartan element, $N_{+},M_{+}$ is upper
triangular and $N_{-},M_{-}$ is lower triangular.

Acting $U_{L}\left( U_{R}\right) $ on the highest (lowest) weight vector of
the level one representation of affine algebra. The upper(lower) triangular $%
N_{+}\left( N_{-}\right) $ factor is annihilated. One can show as \cite{b244}
that the remain left moving $M_{-},K_{-}$ are \textquotedblleft \textbf{%
holomorphic}\textquotedblright , right moving $M_{+},K_{+}$ are
\textquotedblleft \textbf{anti-holomorphic}\textquotedblright\ respectively
\begin{equation}
\partial _{+}M_{-}^{-1}=\partial _{+}K_{-}=0\ ,
\end{equation}%
\begin{equation}
\partial _{-}M_{+}^{-1}=\partial _{-}K_{+}=0\ .
\end{equation}

The $M_{\pm }$ are determined by $K_{\pm }$ \cite{bilalgervais}(Serret
Frenet eq. of chiral embedding \cite{gervais})
\begin{equation}
M_{-}\partial _{-}M_{-}^{-1}=e^{-adK_{-}\mathcal{E}_{-}}\ ,  \label{eq34}
\end{equation}%
\begin{equation}
\partial _{+}M_{+}M_{+}^{-1}=-e^{-adK_{+}\mathcal{E}_{+}}.  \label{eq35}
\end{equation}%
where
\begin{equation}
\mathcal{E}_{+}=\lambda \sum_{i=0}^{r}E_{\alpha _{i}},\quad \mathcal{E}%
_{-}=\lambda ^{-1}\sum_{i=0}^{r}E_{-\alpha _{i}},\quad \mathrm{%
principal\quad gradation},
\end{equation}

Then the various combination of the minors of Wronski determinant \cite%
{belog,gervais} of $K_{\mp }$ will be expressed uniquely by the Cartan field
$\Phi $

\begin{equation}
\Phi =\varphi \cdot H+\eta d+\xi C.  \label{38}
\end{equation}%
where $H$ is the finite (level zero) part of $\mathfrak{h}$ and we will see
that $\phi $ is the affine Toda field and $\eta ,\xi $ are the fields
corresponding to the grade derivative $d$ and the center $C$ in Cartan
subalgebra $\mathfrak{h}$ of $gl(4)^{(1)}$ respectively.

Then it turns out \cite{b244} that the $U$ in (\ref{uu}) turns to be the
\textbf{transfer matrix} satisfied (\ref{au}) with \textbf{Lax connection} $%
A_{\pm }$(\ref{a}) expressed solely by $\Phi $
\begin{equation}
A_{+}=\partial _{+}\Phi +me^{ad\Phi }\mathcal{E}_{+}\ ,  \label{42}
\end{equation}%
\begin{equation}
A_{-}=-\partial _{-}\Phi +me^{-ad\Phi }\mathcal{E}_{-}\ .  \label{43}
\end{equation}%
Here we fix $\mu \nu =m^{2}$, and set $\frac{\mu }{m}=\frac{m}{\nu }%
\longrightarrow \mu $.

In terms of (\ref{42})(\ref{43}) the self consistence condition of $\left( %
\ref{au}\right) ,$
\begin{equation}
\partial _{+}A_{-}-\partial _{-}A_{+}+\left[ A_{+},A_{-}\right] =0\ ,
\label{zero}
\end{equation}%
becomes the \textbf{EOM of conformal affine Toda }$\Phi $.

For simplicity, let us restrict to the $sl(2)^{(1)}$ case , the EOM becomes
that of the conformal Sine-Gordon
\begin{equation}
\partial _{+}\partial _{-}\varphi =m^{2}e^{2\eta }\left( e^{2\varphi
}-e^{-2\varphi }\right) ,  \label{sg1}
\end{equation}%
\begin{equation}
\partial _{+}\partial _{-}\eta =0,  \label{sg2}
\end{equation}%
\begin{equation}
\partial _{+}\partial _{-}\xi =m^{2}e^{2\eta }\left( e^{2\varphi
}+e^{-2\varphi }\right) .  \label{sg3}
\end{equation}

\textbf{Conformal invariance}

This EOM is invariant under left ($f(z_{+}))$ and right $(g(z_{-}))$
independent reparametrization \cite{b244}
\begin{equation}
\tilde{\varphi}(\tilde{z}_{+},\tilde{z}_{-})=\varphi (f(\tilde{z}_{+}),g(%
\tilde{z}_{-}))+\frac{1}{2}\log (f^{\prime }g^{\prime }),
\end{equation}%
\begin{equation}
\tilde{\eta}(\tilde{z}_{+},\tilde{z}_{-})=\eta (f(\tilde{z}_{+}),g(\tilde{z}%
_{-}))+\log (f^{\prime }g^{\prime }),
\end{equation}%
\begin{equation}
\tilde{\xi}(\tilde{z}_{+},\tilde{z}_{-})=\xi (f(\tilde{z}_{+}),g(\tilde{z}%
_{-}))-\frac{1}{4}\log (f^{\prime }g^{\prime }).
\end{equation}%
The contribution of the two fields $\eta $ and $\xi $ in the improved
energy-momentum tensor $\Theta _{\mu \nu }$ is \cite{b244,underwood}
\begin{equation}
(\partial _{\mu }\partial _{\nu }-g_{\mu \nu }\partial ^{2})(\frac{1}{2}\eta
-2\xi ),  \notag
\end{equation}%
so the improved energy momentum tensor will be traceless \cite{underwood}
\begin{equation}
\Theta _{\ \mu }^{\mu }=0.
\end{equation}%
The integral of improve energy momentum tensor in 2-dim, will give a surface
term which is the topological term (the first Chern-class) and equals the
number of soliton. The traceless of energy momentum tensor implies that the
\textbf{conformal invariance }is recovered. The total energy and momentum of
one soliton equals
\begin{equation}
H\pm P=m^{2}\exp (\mp \mu ).
\end{equation}

\section{Soliton solution of affine conformal Toda}

Its soliton solutions can be obtained in three way:

\textbf{A}. Solve the equations of motion (\ref{sg1},\ref{sg2},\ref{sg3}) to
obtain one soliton solution, then use the B$\ddot{\mathrm{a}}$cklund
transformation.

\textbf{B}. By \textbf{Riemann-Hilbert} method in connection with \textbf{%
dressing transformation }\cite{rh}.

The dressing transformation of $U${\ by }${g}${\ to }$U^{g}$ is solved by
the following Riemann-Hilbert method.

Factorize
\begin{equation}
W\left( g\right) =UgU^{-1}
\end{equation}%
into
\begin{equation}
W=W_{-}^{-1}W_{+}
\end{equation}%
with $W_{-},W_{+}$ analytic respectively in two region, e.g. upper and lower
complex plane.

Then set
\begin{equation}
U^{g}=\bigg\{
\begin{array}[l]{l}
W_{+}Ug_{+}^{-1}\quad \mbox{\rm in upper plane} \\
W_{-}Ug_{-}^{-1}\quad \mbox{\rm in lower plane}\ ,%
\end{array}%
\end{equation}
where
\begin{equation}
g^{-1}_- g_+ = g .
\end{equation}
It is easy to check
\begin{equation}
W\left( g\right)
=UgU^{-1}=Ug_{-}^{-1}g_{+}U^{-1}=W_{-}^{-1}U^{g}U^{g-1}W_{+}=W_{-}^{-1}W_{+}
\end{equation}

Consequently the equations
\begin{equation}
\partial _{i}U=A_{i}U
\end{equation}%
is transformed into
\begin{equation}
\partial _{i}U^{g}=A_{i}^{g}U^{g}.
\end{equation}%
Thus R.H transformations $W_{\pm }$ induce different \textquotedblleft gauge
transformation\textquotedblright\ on the Lax connection $A_{i}$ of $U$ to
get the same%
\begin{equation}
A_{i}^{g}=W_{\pm }A_{i}W_{\pm }^{-1}-\partial _{i}W_{\pm }W_{\pm }^{-1},
\end{equation}%
actually it is shown that \cite{dressing}
\begin{equation}
W_{+}=K_{+}^{g}M_{+}^{g}\ ,
\end{equation}%
\begin{equation}
W_{-}=K_{-}^{g}M_{-}^{g}\ ,
\end{equation}%
\begin{equation}
K^{\pm g}=\exp \pm (\Phi -\Phi ^{g})\ .
\end{equation}%
The dressing transformation form a dressing group with the following
multiplication rule
\begin{equation}
g\circ h=(g_{-}h_{-})^{-1}(g_{+}h_{+})\ .
\end{equation}

Here $g$ is decomposed as
\begin{equation}
g=g_{-}^{-1}g_{+}\ ,
\end{equation}%
where $g_{\pm }$ are
\begin{equation}
g_{\pm }=\exp {\hat{H}}\exp {\hat{N}_{\pm }}\subset \pm \text{ Borel subgroup%
}
\end{equation}%
$\hat{H},\hat{N}_{\pm }$ are the factor of the Gaussian decomposition of $g$.

\textbf{C}. By using vertex operator (e.g. principal realization) of
Kac-Moody algebra and factorized it into $V^{\pm}$ in connection with
dressing transformation.

\subsection{One soliton solution}

\textbf{I}. The Riemann-Hilbert method with one zero(pole) at $\pm \mu $
\cite{rh}
\begin{equation}
W_{1}=1-\frac{2\mu }{\lambda +\mu }P,\quad W_{2}=1+\frac{2\mu }{\lambda -\mu
}P,\quad P^{2}=P.  \label{P}
\end{equation}%
can be solved by using the dressing transformation from the vacuum solution
as followings \cite{babelon}.

The vacuum solution
\begin{equation}
\varphi _{vac}=0,\quad \eta _{vac}=0,\quad \xi _{vac}=2m^{2}z_{+}z_{-},
\end{equation}%
\begin{equation}
U_{vac}=e^{\frac{m^{2}}{2}z_{-}z_{+}C}e^{-mz_{-}\mathcal{E}_{-}}e^{-mz_{+}%
\mathcal{E}_{+}}.  \label{70}
\end{equation}

Let \cite{babelon}
\begin{equation}
g_{-}=e^{-2\log \mathrm{cosh\frac{\varphi _{sol}}{2}C}}\tilde{g}_{-}\ .
\label{sol}
\end{equation}%
Then the dressing eq.%
\begin{equation}
U_{sol}(x,t)=g_{\pm }(x,t)U_{vac}(x,t)g_{\pm }^{-1}(0)\ ,  \label{72}
\end{equation}%
turns to be%
\begin{equation*}
\left( \partial _{+}+m\lambda \left(
\begin{array}{cc}
0 & 1 \\
1 & 0%
\end{array}%
\right) \right) \hat{g}_{-}^{-1}=\hat{g}_{-}^{-1}\left(
\begin{array}{cc}
\frac{1}{2}\partial _{+}\varphi _{sol} & m\lambda e^{\varphi _{sol}} \\
m\lambda e^{-\varphi _{sol}} & -\frac{1}{2}\partial _{+}\varphi _{sol}%
\end{array}%
\right)
\end{equation*}%
Its one pole solution is%
\begin{equation}
\tilde{g}_{-}^{-1}=\left(
\begin{array}{cc}
e^{-\frac{\varphi _{sol}}{2}} & 0 \\
& e^{\frac{\varphi _{sol}}{2}}%
\end{array}%
\right) -\mathrm{sinh}\frac{\varphi _{sol}}{2}\left(
\begin{array}{cc}
\mu & \lambda \\
-\lambda & -\mu%
\end{array}%
\right) \frac{\mu }{\lambda ^{2}-\mu ^{2}}\mathrm{\ ,}  \label{drerh}
\end{equation}%
$g_{+}$ similar.

If we set the
\begin{equation}
\frac{1}{\sqrt{\lambda ^{2}-\mu ^{2}}}\left(
\begin{array}{cc}
\mu  & \lambda  \\
-\lambda  & -\mu
\end{array}%
\right) =\frac{\left( \mu \sigma _{3}+\lambda i\sigma _{2}\right) }{\sqrt{%
\lambda ^{2}-\mu ^{2}}}\equiv \sigma _{3}^{\prime }
\end{equation}%
as the direction of the common tangent \cite{chen,hhbook,houo3}, define
\begin{equation}
\sigma _{3}^{\prime }=P+P_{\perp },P\cdot P_{\perp }=0,P^{2}=P.
\end{equation}%
Then we can show that $P$ is the projection operator in the eq.(\ref{P}).

It remains to get the $\varphi _{sol}$ in eq.(\ref{drerh}),\textbf{\ }there
are three ways \cite{babelon}:

1. Solve the affine Toda equation (\ref{sg3}) with $\eta =0$ to get
\begin{equation}
\mathrm{tanh}\frac{{\varphi }_{sol}}{2}=-a\exp [2m(\mu z_{+}+\mu
^{-1}z_{-})];  \label{ssolu}
\end{equation}

2. Further solve the factorized dressing eq.(\ref{72}) by substitute the $%
U_{vac}$ (\ref{70}) into it to obtain%
\begin{equation}
g_{-}^{-1}(x,t)g_{+}(x,t)=e^{-mz_{-}\mathcal{E}_{-}}e^{-mz_{+}\mathcal{E}%
_{+}}g_{-}^{-1}(0)g_{+}(0)e^{mz_{+}\mathcal{E}_{+}}e^{mz_{-}\mathcal{E}%
_{-}}\ .  \label{bb62}
\end{equation}%
Then solve this eq. as in $\S $ 5.1 of ref.\cite{babelon};

3. The most elegant method is to use the factorized vertex operators as in
\textbf{II}.

\textbf{II}. The g written in exponential forms becomes%
\begin{eqnarray}
\tilde{g}_{-}^{-1}\left( x,t\right)  &=&g_{-}^{-1}e^{2\log \mathrm{cosh\frac{%
\varphi }{2}C}}=e^{-\varphi _{sol}[\frac{1}{2}H+\frac{(\frac{\lambda }{\mu }%
)^{-2}}{1-\left( \frac{\lambda }{\mu }\right) ^{-2}}H+\frac{(\frac{\lambda }{%
\mu })^{-1}}{1-\left( \frac{\lambda }{\mu }\right) ^{-2}}\left(
E_{+}-E_{-}\right) ]}\   \label{h1} \\
&\equiv &e^{-\phi _{sol}V_{\mu }^{\left( -\right) }},
\end{eqnarray}%
\begin{equation*}
\tilde{g}_{+}=e^{\phi _{sol}V_{\mu }^{\left( +\right) }}.
\end{equation*}%
Here
\begin{equation}
V_{\mu }^{(\pm )}=\frac{1}{2}H+\frac{(\frac{\lambda }{\mu })^{\pm 2}}{%
1-\left( \frac{\lambda }{\mu }\right) ^{\pm 2}}H+\frac{(\frac{\lambda }{\mu }%
)^{\pm 1}}{1-\left( \frac{\lambda }{\mu }\right) ^{\pm 2}}\left(
E_{+}-E_{-}\right) ,  \label{h3}
\end{equation}%
are the positive and negative frequency respectively of $V(\lambda ),$
\begin{equation}
V_{\mu }^{(+)}=\frac{1}{2}{\oint_{|\lambda |<|\mu |}}\frac{d\lambda }{2\pi
i\lambda }\frac{\lambda +\mu }{\lambda -\mu }V(\lambda )\ ,
\end{equation}%
\begin{equation}
V_{\mu }^{(-)}=\frac{1}{2}\oint_{|\lambda |>|\mu |}\frac{d\lambda }{2\pi
i\lambda }\frac{\lambda +\mu }{\lambda -\mu }V(\lambda )\ ,
\end{equation}%
where $\mu $ is the position of pole \cite{kac} and $\lambda $ is the
spectral parameter. The the vertex operators principal construction $%
V(\lambda )$ is defined by
\begin{equation}
V(\lambda )=\frac{1}{2}:\exp (-2\sum_{n\ odd}a_{-n}\frac{\lambda ^{n}}{n}):
\end{equation}%
where oscillators $a_{m}$ satisfy
\begin{equation}
\left[ a_{m},a_{n}\right] =m\delta _{m+n,0}\quad \mbox{for m, n
odd},
\end{equation}%
\begin{equation}
a_{m}|0\rangle =0,n>0\ .
\end{equation}

Then using
\begin{equation}
\left[ E_{+}^{\left( n\right) }+E_{-}^{\left( n\right) },V(\lambda )\right]
=-2\lambda ^{n}V(\lambda ),  \label{e1}
\end{equation}%
here and in (\ref{h1}-\ref{h3}),%
\begin{eqnarray*}
E_{\pm }^{\left( 2n-1\right) } &=&\lambda ^{2n-1}E_{\pm },H^{\left(
2n\right) }=\lambda ^{2n}H\qquad n>0, \\
E_{\pm }^{\left( 2n+1\right) } &=&\lambda ^{2n+1}E_{\pm },H^{\left(
2n\right) }=\lambda ^{2n}H\qquad n<0.
\end{eqnarray*}%
It is easy to show%
\begin{equation}
e^{-mz_{-}\mathcal{E}_{-}}e^{-mz_{+}\mathcal{E}_{+}}V(\mu _{i})e^{mz_{+}%
\mathcal{E}_{+}}e^{mz_{-}\mathcal{E}_{-}}=e^{2m(\mu _{i}z_{+}+\mu
_{i}^{-1}z_{-})}V(\mu _{i})  \label{e2}
\end{equation}%
compare it with the eq.(\ref{bb62},\ref{h1}), then one find the $\varphi
_{sol}$ is the same as (\ref{ssolu}).

\subsection{N-solitons solution}

Similarly the n-soliton solutions of conformal affine model as the orbit of
the vacuum solution under the dressing group may be construct from dressing
product of one soliton solution. It is easy to see from (\ref{ssolu}),(\ref%
{h1}) that
\begin{equation*}
g_{\pm }\left( 0\right) =g_{\pm }\left( x,t\right) |_{x=0,t=0}
\end{equation*}%
may be expressed by $1+aV\left( \mu \right) .$ Then as shown by Ref.\cite%
{babelon} their dressing product equals
\begin{equation}
g\left( 0\right) =\left( 1+2a_{1}V(\mu _{1})\right) (1+2a_{2}V(\mu
_{2}))\cdots (1+2a_{N}V(\mu _{N})).
\end{equation}%
where $a_{i,}\mu _{i}$ will be the zero mode position and momentum of the $i$%
-th solitons. So from (\ref{bb62}) the $N$-solitons solutions may be
expressed by%
\begin{equation*}
\left\langle \Lambda ^{\pm }\left\vert \prod\limits_{i=1}^{n}\left(
1+2a_{l}\exp \left[ 2m(\mu _{l}z_{+}+\mu _{l}^{-1}z_{-})\right] V(\mu
_{l})\right) \right\vert \Lambda ^{\pm }\right\rangle
\end{equation*}%
Further by factorize each normal ordered $V\left( \mu \right) $ into $%
V_{+}\left( \mu \right) $ and $V_{-}\left( \mu \right) ,$ shift $V_{\pm }$
to left or right, then by Wick theorem, the usual $\tau $ function will be
obtained.

\subsection{Soliton of chiral embedded string in $sl(4)^{(1)}$}

To find it from the known $sl(2)^{(1)}$ solution, we simply embed the $%
sl(2)^{(1)}$ in the $sl(4)^{(1)}$, the generator $E_{\pm }$ map to the
cyclic elements $\sum\limits_{i=0}^{3}e_{\pm \alpha _{i}}$ of the $sl(4)$.
Since this cyclic element is invariant under the action of the Coxeter
element $\sigma $ which is the maximal Weyl reflection for the automorphism
of the Dynkin diagram \cite{kac}, then it is easy to find that the one
soliton solution of the affine $sl(4)^{(1)}$ is given by \cite{underwood}
\begin{equation}
\tanh {\frac{\varphi _{sol}^{i}}{2}}=A\exp \left[ m(\omega ^{i-1}\mu
z_{+}-\omega ^{1-i}\mu ^{-1}z_{-})\right] ,\quad i=0,\cdots ,3.
\end{equation}%
where
\begin{equation}
\omega =e^{\frac{2\pi i}{h}},
\end{equation}%
here coxeter number $h$ equals $4$. The n-soliton solutions of $sl(4)^{(1)}$
are similar with \textbf{5.2}.

\section{The stretched string soliton}

Since the $AdS_{5}$ space has the Poincare metric in conformal flat
coordinate. We illustrate the motion of string in $AdS_{5}$ by the following
figures

\begin{figure}[H]
\begin{center}
\includegraphics{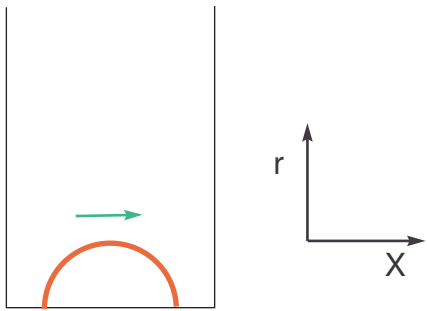}
\end{center}
\caption{In the conformal flat Poincare coordinate of $AdS_{5}$, The two
sides of upper half \textquotedblleft strip\textquotedblright\ is
identified. $r$ is the inverse radius of $S^{5}$ and $X$ is the other four
coordinates. The semicircle is the $\protect\sigma $\ orbit of the stretched
string at the equal time $\protect\tau $. The string will move to right
(left) around the half cylinder, or on its universal enveloping upper
\textquotedblleft plane\textquotedblright .}
\end{figure}

\begin{figure}[H]
\begin{center}
\includegraphics{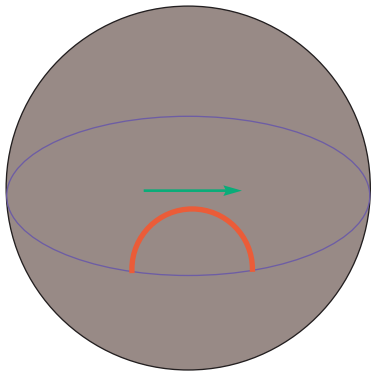}
\end{center}
\caption{The deform of string in $AdS_{5}$ (wick rotated to $S^{5}$). The
equator is the horizon and the string approach it. With time evolving the
left (right) moving string move anti-clockwise (clockwise).}
\end{figure}

The metric of $AdS_{5}$ in the pure bosonic part of action $S_{k}$ $\left( %
\ref{sk}\right) ,$by (\ref{11}), becomes the Poincare metric
\begin{equation}
\frac{\sum_{i=1}^{4}dx_{i}^{2}+dr^{2}}{r},
\end{equation}%
for the pseudo-sphere. The world-sheet $\sigma $ and $\tau $ maps to the two
principal curvature coordinates. The normal image of the motion on
pseudo-sphere will describe the dynamics of nonlinear $\sigma $ model \cite%
{houo3}, now on $AdS^{5}$.

The ground states of solitons is the twistly stretched string on $AdS_{5}$.
The two end points of string approach the horizon and the left (right)
moving string at fixed time $\tau $ is described by a geodesic line on the
pseudo-sphere, which is the semicircle in fig.1 and fig.2. The geodesic
radius of the semicircle equals the exponential rapidity $\mu $ and it
evolves with constant velocity in time counterclockwise (clockwise) around
with period $\frac{1}{2\pi }$.

\section{Twistor Description}

We have shown that the IIB Green-Schwarz superstring in $AdS_{5}\times S^{5}$
is described by the conformal $su(4)^{(1)}$ affine Toda theory. This affine
algebra $su(4)^{(1)}$ which is the affinization of the conformal group $%
SU(4)\sim O(4,2)$ of the coset space $AdS_{5}$. The conformal group acts on
the twistor space $\mathbb{T}\sim \mathbb{CP}^{3}$, as implies by \cite{RS}
and \cite{witten1}.

Let us as Maldacena \cite{mal} denote the coordinates $(t,v,w,x,y,z)$ of a
point in $AdS_{5}$
\begin{equation}
t^{2}+v^{2}-w^{2}-x^{2}-y^{2}-z^{2}=R^{2}.
\end{equation}%
The metric of $AdS_{5}$ turn to be
\begin{equation}
ds^{2}=dt^{2}+dv^{2}-dw^{2}-dx^{2}-dy^{2}-dz^{2}.
\end{equation}

But a point in $AdS_{5}$ correspond to a bilinear form $\mathsf{\ R}^{ab}$
\cite{penrose} i.e. a line in $\mathbb{T}$
\begin{equation}
\mathsf{R}^{ab}=Z^{a}X^{b}-X^{a}Z^{b},\quad Z^{a}=(\lambda ^{\alpha },\mu _{%
\dot{\alpha}})\ ,
\end{equation}%
\begin{equation}
\mathsf{R}^{01}=\frac{1}{2}\left( v+w\right) ,\mathsf{R}^{02}=\frac{1}{\sqrt{%
2}}\left( y-ix\right) ,\mathsf{R}^{03}=\frac{i}{\sqrt{2}}\left( t+z\right) ,
\end{equation}%
\begin{equation}
\mathsf{R}^{12}=\frac{i}{\sqrt{2}}\left( z-t\right) ,\mathsf{R}^{13}=\frac{1%
}{\sqrt{2}}\left( y+ix\right) ,\mathsf{R}^{23}=v-w,
\end{equation}%
here it is not a linear rotation of $SU(4)\sim SO(4,2)$ in $AdS_{5}.$

The $SU(4)\sim SO(4,2)$ generators are rotations in $\mathbb{T}$\ \cite%
{witten1}
\begin{equation}
e^{\left[ ab\right] }=Z^{a}\frac{\partial }{\partial \bar{Z}_{b}}-Z^{b}\frac{%
\partial }{\partial \bar{Z}_{a}},
\end{equation}%
\begin{equation}
\bar{Z}_{a}=(\bar{\mu}_{\alpha },\bar{\lambda}^{\dot{\alpha}}).
\end{equation}%
Now we affinize it into $\hat{e}^{[ab]}$ of $SU(4)^{(1)}$ in homogeneous
gradation. Let
\begin{equation}
\hat{e}^{aa+1}\equiv \hat{Z}^{a}\frac{\partial }{\partial \hat{\bar{Z}}_{a+1}%
}=\lambda e^{aa+1}=\lambda Z^{a}\frac{\partial }{\partial \bar{Z}_{a+1}}%
,\quad a\in \mathbb{Z}_{4}\ .
\end{equation}

Subsequently, the original real point $x_{\alpha \dot{\beta}}$ in $\mathbb{M}%
^{4}$
\begin{equation}
\lambda ^{\alpha }x_{\alpha \dot{\beta}}=\mu _{\dot{\beta}}
\end{equation}%
becomes a complexified point $\hat{x}_{\alpha \dot{\beta}}$ in $\mathbb{CM}%
^{4}$
\begin{equation}
\hat{\lambda}^{\alpha }\hat{x}_{\alpha \dot{\beta}}=\hat{\mu}_{\dot{\beta}}\
.
\end{equation}

The motion of IIB string will be described by
\begin{equation}
\bar{Z}_{\alpha }Z^{a}=m^{2}=2s\neq 0.  \label{100}
\end{equation}%
and the Robinson congruence $X^{a}$ as following.

The Robinson congruence $X^{a}$ intersecting $Z^{a}$ satisfy
\begin{equation}
X^{a}\bar{Z}_{a}=0
\end{equation}%
\begin{equation}
X^{a}\bar{X}_{a}=0  \label{rc}
\end{equation}%
Let
\begin{equation}
X^{a}=(\lambda ,-s,\mu ,1),\qquad \bar{X}_{a}=(\bar{\mu},1,\bar{\lambda}%
,-s)\ ,
\end{equation}%
then from eq.(\ref{rc}) we have
\begin{equation}
Re(\lambda \bar{\mu})=s.
\end{equation}%
The $\frac{\lambda }{\sqrt{s}}=\frac{\sqrt{s}}{\mu }$ will correspond to our
loop parameter $\mu =\frac{1}{\nu }$.

As well known, the real points in $\mathbb{CM}^{4}$\ Klein correspond to
lines of $\mathbb{PT}$ that lie entirely in the subspace $\mathbb{PN}$ of
null projective twistors $Z^{a},$%
\begin{equation*}
Z^{a}\bar{Z}^{a}=0.
\end{equation*}%
That is, $Z^{a}$ intersects (lies on) its conjugate plane. This null $Z^{a}$
correspond to geodesic null congruence of null lines in $\mathbb{CM}^{4},$
the distance between any two null lines in this congruence is zero. Now the
Robinson congruence $X^{a}$ intersecting non-null $Z^{a}$ (\ref{100}-\ref{rc}%
) is not geodesic null, it remains to be shear free, while $|\mu |$ (and its
phase) correspond to the nonzero expansion (and rotation) spin coefficient
respectively. For the ray congruence, confer Trautman and Pirani \cite{prani}
and Sachs \cite{sachs}. Later maybe we'll give the detail of these
descriptions and the relation between this twistor congruence of string with
the pseudo congruence \cite{chen} and with the Backlund transformation and
the dual twist rotations on pseudo-sphere, but the more urgent is to
describe the physical properties of stretched string, e.g. its
\textquotedblleft electromagnetic\textquotedblright\ behavior. All this will
be more transparent, if we consider the move of 4-bein frame of twistor, the
\textquotedblleft flag\textquotedblright , which describe the
(anti)self-dual YM at the same time. So we'll study the super YM case in
next paper.

\section{Discussion}

Since the WZW term of RS action (\ref{eq8}) contains solely
fermionic current, we cannot constrain it directly to give the
pure bosonic gauged WZW action \cite{mtsu4}, but have to introduce
the super current version of (\ref{eq8}), then constrain it as a
gauged WZW action and reduce to bosonic (\ref{acction}). This way
of formulation is not satisfactory. The better would be an
superspace action generalize that of, e.g. chiral super gravity
\cite{west}. There should be two covariant
constant super field, one is the superfield of the normal vector $Y$ in Ref.%
\cite{RS}, the another describes the $U(1)$ gauge in S duality $SL(2,R)$.

In the superspace formulation the constraint \cite{awitten} turns
to be chiral covariance of e.g. the $N=1$ vector multiplet $V$ and
chiral multiplet $\Phi ,$ both from the $N=2$ vector multiplet.
The chiral anomaly combine the contributions both in the D term
and in the superpotential into e.g. the Chern Simons term
\cite{moore}. The dilation and phase twist of VEV determines the
K\"{a}hler twist and complex twist \cite{twist} in chiral ring and
special geometry \cite{specialg}.

We are more interested to investigate the dressing symmetry of twisted dual
more explicitly in these topics and to find the role played by the twistly
stretched string and twisted monopole to resolve the singularity, to blow up
in topological transition \cite{wvafa}.

As a preliminary, in next paper \textquotedblleft The affine ambitwistor
space as the moduli space of SUYM in $AdS_{5}\otimes S^{5}$%
\textquotedblright , we will study the more realistic case, SDYM in 4d space
time. There, as is well known in Sieberg Witten's paper on monopole
condensation \cite{sw} the spectral curve is given by the spectral
determinant of affine Toda. There should be some integrable model \cite%
{dwitten}, but why it is just the affine Toda has been consider as
mysterious \cite{marshakov}. It is plausible that the dressing symmetry
dictate uniquely that the spectral curve will be from the affine Toda.

\section*{Acknowledgments}

We thank Xiang-Mao Ding, Kang-Jie Shi and Wen-Li Yang for their helpful
discussions.

\section*{Note}

The left (right) moving world-sheet embedded respectively into
wave functions $ |\xi>$ ($ <\xi|$) of the two N=2 fermions with
opposite chirality. The wave function $ |\xi> \subset $ twistor
space $CP_{3}\sim$ fundamental representation of $ U(4)\sim
O(4,2)$, which are central extended into the affine $
\widehat{U(4)}$. They satisfy the Killing spinor equation, which
are the $ \delta\psi_{gaugino}=\delta\psi_{gravitino}=0$, dual
twisted by the phase of the central charge of the topological
gauge field. Meanwhile the rapidity $ \mu$ assumes the attractor
limit value.  The $ |\xi>$  and $ <\xi|$ instead of being
conjugate pure states, are mixed states with density matrix,
realizing the dilation boost. Thus the holomorphic anomaly of
almost factorization is given by the central term of the affine
vertex, i.e. the soliton generating operator.

\end{document}